\documentclass{aa}

\begin{document}

\thesaurus{02.01.2; 02.16.1; 06.06.3; 11.01.2; 13.25.3  }
\title{Effect of Beam-Plasma Instabilities on 
Accretion Disk Flares}

\author{Vinod Krishan\inst{1}
\and Paul J. Wiita\inst{2}
\and S. Ramadurai\inst{3} }

\offprints{V.\ Krishan}

\institute{
Indian Institute of Astrophysics, Koramangala, Bangalore 560034, 
India\\
email: vinod@iiap.ernet.in
\and
Georgia State University, Department of Physics and Astronomy,
Atlanta GA, 30303, USA\\
email: wiita@chara.gsu.edu
\and
Tata Institute of Fundamental Research, Theoretical
Astrophysics Group, Homi Bhabha Marg, Mumbai 400005, India\\
email: durai@tifr.res.in}

\date{Received .../ Accepted ....}

\authorrunning{Krishan et al.}
\titlerunning{Beam-plasma instabilities}
\maketitle

\begin{abstract}
We show that a certain class of flare models for variability from accretion
disk coronae are subject to beam-plasma instabilities.  These instabilities
can prevent significant direct acceleration and greatly reduce
the variable X-ray emission argued to arise via inverse Compton
scattering involving relativistic electrons in beams and soft
photons from the disk.

\end{abstract}
 
\section{Introduction}

The origin of fluctuations in the emission from Active Galactic Nuclei
(AGN) and binary X-ray sources is an important and long-standing problem.
One frequently considered  possibility employs flares in the coronae around
accretion disks to produce rapid energy release, particle acceleration
and radiation (e.g., Galeev, Rosner \& Vaiana \cite{galeev}; Kuperus 
\& Ionson \cite{kuperus};
for a review, see Kuijpers \cite{kuijpers}).
These models usually build upon our understanding of solar flare physics.

A particular model of this type has been proposed by de Vries \& Kuijpers
(\cite{deVries}; hereafter dVK), and was specifically applied by them to 
X-ray variability of
AGNs.  Their model is an elaboration on typical flare scenarios,
in that, as usual,  the source of energy is stored in magnetic fields
 in coronae
of accretion disks.  They estimate the power released
in flares in a radiation pressure dominated corona, which they stress
is a different environment from the gas pressure dominated solar corona. 
They argue this leads to a situation where beams of relativistic
electrons are produced in the corona and then lose essentially all
of their energy through
inverse Compton scattering on UV disk photons before they can stream
back to the 
disk.  They further argue that these inverse Compton (IC) photons produce
the X-ray variability seen in Seyfert galaxies, and are able to calculate
spectral power-densities in reasonable agreement with observations.

However, the dVK  model does not take into account other mechanisms
that might vitiate some of their key assumptions.  We note that dVK  
briefly argue that, particularly if radiation pressure dominates the
energy density in the corona, as is indeed likely around standard thin
accretion disks (e.g., Shakura \& Sunyaev \cite{shakura})
which they assume, energy losses through scattering on
plasma waves are unimportant; then
the dominant losses will be to IC scattering.  However,
it is well known that an electron beam-plasma system is often susceptible
to the excitation of beam-plasma instabilities which usually have
large growth rates (Sturrock \cite{sturrock}).  Here we argue that
when these beam-plasma instabilities (BPIs) are
taken into account, the rate of loss of energy by the electrons 
for the accretion disk coronae conditions suggested by dVK is
typically much higher than the rate of gain of energy through 
direct acceleration by the electric fields, which are
presumed to arise in reconnection events.
Therefore beams of electrons usually will not  reach the high Lorentz factors
needed to produce most X-rays by the IC process.  In many accretion disk 
models X-rays are usually produced through IC scattering of soft photons on
hot thermal electrons (e.g. Shapiro, Lightman \& Eardley \cite{shapiro}; Liang \&
Price \cite{liang}).
In such a situation beam-plasma instabilities are not excited,
and  only thermal spontaneously excited plasma waves should exist.  These
will have energy
densities less than the thermal energy density of the plasma, which in
turn is much less than the radiation energy density.  In this case,
the argument of  dVK would be valid, but, once they assume a beam is
present, then beam-plasma instability effects {\it must} be included. 

\section{Growth of Beam-Plasma Instabilities}

The key assumptions of the dVK model are that: 1) relaxation of magnetic
structures efficiently produce relativistic electron beams; 
2) the particle beam is a mono-energetic
stream of electrons with an initial Lorentz factor $\gamma_0$; 3) the ambient
radiation is from a quasi-infinite disk and can be considered as uniform and
isotropic, with a radiation density $u_{\rm rad}$; 4) the beam is optically
thin, so multiple scattering of photons can be ignored.  Although (3) is an
approximation, it is a reasonable one, and (4) is certainly plausible under
many circumstances.  But the core of their argument hinges on the ability of 
the neutral sheet in the reconnection process to quickly accelerate electrons
via a direct electric field.  During this acceleration process dVK claim 
the equation for the acceleration of an single electron suffering
IC losses is                        
\begin{equation}                
{\frac {d \gamma}{dt}} = \chi_1 {\frac {(\gamma^2 - 1)^{1/2}}{\gamma}}
   - \chi_2 (\gamma^2 - 1),
\end{equation}
where, $\chi_1 = eE/m_e c$ and $\chi_2 = 4\sigma_T u_{rad} / 3m_e c$,
with all symbols having their usual meanings.
In that the first (positive) term starts out substantially greater in
magnitude than the second (negative) one, acceleration will ensue until  
a limiting Lorentz factor is reached when the two terms balance: 
\begin{equation}
\gamma_{\infty} = 2^{-1/2}[1 + (1 + 4 \chi_1^2 / \chi_2^2)^{1/2}]^{1/2}.
\end{equation}

The electric field is reasonably taken by dVK to be the Dreicer value, which
we take as: $E_D = 6 \pi n_p e^3  {\rm ln} \Lambda /(k_B T_e)$, where $n_p$ is
the electron density of the ambient plasma, ${\rm ln} \Lambda \approx 20$
is the Coulomb logarithm, and all other symbols have
their usual meanings.
With typical AGN values ($n_p \approx 10^{10}$ cm$^{-3}$, $T_e \approx 10^6$K,
and $T_{rad} \approx 10^5$K) they find $\gamma_{\infty} \approx
(\chi_1/ \chi_2)^{1/2} \approx 30$.  They then conclude that the electrons will
all reach this terminal Lorentz factor  before the acceleration terminates
and the electrons then lose their energy against
the disk photons providing the background radiation field.

We now show that since  a BPI is excited, it will 
dominate the energy losses for the beam and actually
prevent the electrons from reaching the high Lorentz factors calculated
by dVK.
Under these circumstances there will be very little IC radiation, so that,
while a great deal of energy may be released through magnetic reconnection,
the bulk of the energy will probably provide heating to the corona (e.g.,
Liang \& Price \cite{liang}) but is unlikely to yield the bulk of the
X-rays directly through IC emission.

The dominant growth rate of the BPI depends on
the relative magnitudes of the bulk velocity of the beam, $v_b$,  and the 
mean thermal velocity in the beam, $v_{Tb}$; under some conditions,
$v_{Te}$, the mean thermal velocity of the ambient electrons, also must
be taken into account.
The standard formula for the BPI growth rate, valid for 
$v_b > (n_p/n_b)^{1/3} v_{Tb}$, is our Case 1 (e.g., Mikhailovskii 
\cite{mikhailovskii})  
\begin{equation}
\Gamma_{bp} = 0.7 \left({\frac {n_b}{n_p}}\right)^{1/3} \omega_{pe},
\end{equation}
where $n_b$ is the beam density, $n_p$ is the ambient plasma density
(here, in the disk corona),
and $\omega_{pe}= 5.47 \times 10^4 n_e^{1/2}$ is the plasma frequency in terms
of the ambient electron number density in cgs units.   The frequency at 
which this mode grows is $\omega_{pe}(1-0.4(n_b/n_p)^{1/3})$.  

If the beam starts out very slowly, with 
$v_b < (n_p/n_b)^{1/3} v_{Tb}$, then
the ``weak'' version of the BPI is relevant, and this is
our Case 2 (e.g., Benz \cite{benz})    
\begin{equation}
\Gamma_{bp,w} = \left({\frac {n_b}{2n_p}}\right) 
                \left({\frac {v_b}{v_{Te}}}
			\right)^2 \omega_{pe},
\end{equation}
and the frequency at which this dominant mode is excited is $\omega_{pe}$.
Under the limited circumstances that $v_{Te} > v_b > v_{Tb}$, the
``hot-electron'' Case 3 yields (e.g., Mikhailovskii \cite{mikhailovskii}),    
\begin{equation}
\Gamma_{bp,he} = \left({\frac {n_b} {n_p}}\right)^{1/2}
                {\frac{v_{Te}}{v_b}}
		\omega_{pe},
\end{equation}
where this dominant mode is at a frequency of $(v_b / v_{Te}) \omega_{pe}$.

The AGN corona values of dVK for $n_p = 10^{10}$ cm$^{-3}$ and $T = 10^6$ K, 
which we also believe are reasonable,  will
be adopted here.  There are, however,  additional parameters that 
must be considered
now (basically in lieu of the radiation temperature, or $u_{rad}$,
needed by dVK).  First, 
$\zeta \equiv n_b/n_p$;
for solar flares this value is $\sim 10^{-6}$ -- $10^{-4}$ (Benz \cite{benz}); 
however, we will bear in mind the possibility that this ratio may be 
higher in this type of radiation dominated plasma.  
We also need initial values of $v_b$ and $v_{Tb}$, to
determine which of the three Cases defined above should be considered.
For us to say that a beam actually exists we must always
demand that $v_b > v_{Tb}$.  

Note that the BPI directly gives the  rate of growth of an electric 
field in the plasma, and the energy loss goes as the square of the field
strength. Then  we find that when the relativistic effects that arise if 
the Lorentz factors really could become large are included,  the rate of
change of energy of electrons in the beam is,     
\begin{equation}
{\frac {d\gamma}{dt}} = \chi_1 {\frac{(\gamma^2 - 1)^{1/2}}{\gamma}} - 
                        2 \alpha \gamma \Gamma_{bp}(\gamma),
\end{equation}
where $\alpha \equiv W/E = W/\gamma n_b m c^2$, is the ratio between
the wave energy density, $W$, and the electron beam energy density, $E$.
In order to determine $W$, knowledge of the saturation mechanisms of the wave
 field are needed. Often, in order to avoid a detailed discussion of the saturation
mechanisms, which tend to operate in multiplicity in a plasma, the
condition of equipartition of energy between the waves and the beam
particles is used (Treumann \& Baumjohann \cite{treumann}).  In that case 
$\alpha$ is
approximated to unity, and we consider this situation first.  
Case 4, where $\alpha \ll 1$, and the saturation
occurs earlier by trapping, will then be addressed.

In Eqn.\ (6) we have ignored the IC term appearing in Eqn.\ (1), having replaced it with
a generic form of the BPI growth rate; the fact that the BPI term is much
bigger than  the $\chi_2$ term for all reasonable circumstances 
 will soon become evident. 
The dominant dependence of $\Gamma_{bp}$ 
upon $\gamma$ for the first three cases arises through
the replacement:  $n_b \longrightarrow n_b/\gamma^3$ (e.g., Walsh \cite{walsh};
Krishan 1999), which effectively
modifies $\zeta$, which is defined as the density ratio at non-relativistic
 relative velocities.  In Cases
(2) and (3) we must also write $v_b/c = (\gamma^2 - 1)^{1/2}/\gamma$. 

Now, for Cases 1, 2 and 3, respectively, we have:
\begin{equation}
{\frac {d\gamma}{dt}} = \chi_1{\frac{(\gamma^2 -1)^{1/2}}{\gamma}} 
			- 2 A_1,
\end{equation}
with $A_1 = 0.7 \zeta^{1/3} \omega_{pe} \simeq 
8 \times 10^7 \zeta_{-5}^{1/3} n_{e,10}^{1/2}$,
where the common notation, $X_{n} = X/10^{n}$, has been employed
so that the physical parameters will be of order unity; 
\begin{equation}
{\frac {d\gamma}{dt}} = \chi_1{\frac{(\gamma^2 -1)^{1/2}}{\gamma}}
		- 2{\frac{(\gamma^2 -1)}{\gamma^4}}A_2,
\end{equation}
with $A_2 = 0.5 (c/v_{Tb})^2 \zeta \omega_{pe} \simeq 3 \times 10^8
\eta_{b,-2}^{-2} \zeta_{-5} n_{e,10}^{1/2}$, where we have now defined
$\eta_b \equiv v_{Tb}/c \sim 0.01$;
\begin{equation}
{\frac {d\gamma}{dt}} = \chi_1{\frac{(\gamma^2 -1)^{1/2}}{\gamma}} 
		-{\frac {2 \gamma A_3}{(\gamma^3 - \gamma)^{1/2}}},
\end{equation}
with $A_3 = \zeta^{1/2}(v_{Te}/c) \omega_{pe} \simeq 2 \times 10^5
\zeta_{-5}^{1/2} \eta_{e,-2} n_{e,10}^{1/2}$, where now,
$\eta_e \equiv v_{Te}/c \sim 0.01$.
 
Under any of these situations we have  $\chi_1 = 5.2 \times 10^1 n_{e,10}
 T_{e,6}^{-1}$
with our definition of $E_D$ (which is slightly larger than that of
dVK, thereby only strengthening our argument).  For any plausible 
initial value of $\gamma \sim 1$ the different
dependences of Eqns. (7--9) upon $\gamma$ are not
important.  What is important is that  $A_1, A_2, A_3 \gg \chi_1 \gg \chi_2$;
i.e., the energy loss term arising from {\it any} form of the BPI completely
dominates over the energy gain term from direct electric field
acceleration.  

We now consider Case 4, where equipartition is not established.   
Under these circumstances, the growth of the Langmuir waves for
the fastest initial beam situation, Case 1, is arrested
by the trapping of the beam electrons.  In this case, the 
ratio $\alpha$ is eventually
given by the saturated value (Melrose \cite{melrose}; Krishan \cite{krishan}),
$\alpha= 9/2 [n_b/(2 n_p \gamma^3)]^{2/3}$, and it can be a rather
small number that reduces the loss rate significantly.  This gives
a chance for the situation envisioned by dVK to occur.
In addition, $\alpha$ initially can start out below the 
saturation value as it arises from thermal fluctuations, and thus
it could allow an
initial thermal runaway.  The detailed spatial and temporal
structure of the reconnection sites will determine if this initial
acceleration can play a significant role.

In spite of these uncertainties, we can  obtain a reasonable 
estimate of the influence of BPI in the situation 
where equipartition is not established.  We again consider all 
three cases discussed above, but we now include electron trapping 
and assume $\alpha$ to take the saturation value. Here the competition
between the IC losses represented by $\chi_2$ and 
the BPI losses
represented by the Melrose $\alpha$ has to be considered carefully.

The inverse Compton term
increases with $\gamma$ whereas the $\alpha$ factor modifying
the BPI term decreases 
with $\gamma$. Thus
demanding that the BPI term is smaller than the IC 
term fixes the
minimum value of $\gamma$ necessary to validate the dVK proposal. 
A detailed
calculation yields the results for the three cases as follows: case (1a),
$\gamma_{min} = 54$; case (2a), $\gamma_{min} = 21$; case (3a),
$\gamma_{min} = 9.16$. Thus it is clear that only in case (3a), 
is it likely that the IC term dominates and hence the dVK 
proposal is valid. This requires
rather special conditions for the flare models to work.

This type of runaway acceleration has been observed in the laboratory 
under specific circumstances which lead to a very weak beam plasma 
instability. 
In laboratory experiments, the runaway electrons are observed detached
from the main body of the plasma, as for
example in a stellarator.  If the runaway electrons hit the tungsten
aperture, they generate X-rays which can be detected.  Provided the
conditions are right, the runaway electrons undergo instabilities
producing plasma oscillations which then couple to the ions. This
principle is applied in the design of some electron tube
oscillators (Rose \& Clark \cite{rose}).  
Thus the runaway electrons can stably propagate under certain 
circumstances, but will be affected by a BPI  if they
do satisfy the conditions for it. These conditions are essentially on
the velocity of the beam and its thermal spread, as we have already
discussed for the first three cases above.

 Often it is found
that a regime of   strong Langmuir waves is quickly reached and these
waves
are further subjected to modulational instabilities. Thus, different
saturation mechanisms operate at different stages of the development of
the instability, depending on beam plasma parameters.  However, under the
circumstances and parameters proposed by dVK, the damping is severe.

\section{Discussion and Conclusions}

We thus conclude that the mechanism proposed by dVK
should not generally work unless much greater densities are possible in the
coronae at the same time that the temperatures are lower, since $\chi_1$
rises faster with $n_p$ than does any form of $\Gamma_{bp}$,
and declines faster with temperature. While denser
coronae should be available around the accretion disks in X-ray binaries,
the ambient temperatures will also be a good deal higher, so we cannot
suggest a physically interesting situation where the BPIs do not dominate.
If one could somehow begin with very large $\gamma$ values, then
the growth rate of the beam-plasma instabilities are reduced.
For Case 1 this does not help, and no solutions for large $\gamma$
are possible; however, for Cases 2 and 3, the relativistic decreases in the
BPI rates are so substantial that high asymptotic $\gamma$ values
are allowed.  This is also true for Case 4, where the saturation
reduces the effectiveness of the BPI; however, even then the
BPI can prevent much acceleration unless the beam already starts
with a substantial value of $\gamma$ or has such a low density
in comparison to the ambient medium that it could not carry
significant power.
Moreover, we see no way to achieve these initially
high $\gamma$ values: that is what the dVK mechanism was supposed
to accomplish, but now appears to be incapable of achieving.

Filamentation, which could produce denser beam fragments, could play a role
by raising $\zeta$ locally.  If any analogy can be drawn with solar flares,
then the presence of rapid irregularities within the Type 3 radio bursts
strongly indicates that the flux tubes are filamentary during the acceleration
phase (e.g., Vlahos \& Raoult \cite{vlahos}).  However, this possibility
is still insufficient to salvage this mechanism for AGN coronae, since even
with $\zeta \sim 1$ the ratios of $A_{1,2,3}/\chi_1 > 1$.  In Case 4,
 where saturation is important in principle, the large value of $\zeta$
implies that
$\alpha \sim 1$ too (for initial $\gamma \sim 1$) so the loss term
still would dominate.

Nonetheless, even with much of the energy going into wave turbulence,
as we have argued, significant IC emission can be possible. 
 This is because (as pointed
out by the referee) trapping and other nonlinear effects can roughly
heat the electrons up to $kT_e \sim e \phi$, with $\phi$ the
electrostatic amplitude of the waves.  Since the energy gain
term (the first on the RHS of Eq. [1]) is essentially a constant,
these `thermalized/trapped' electrons can attain nearly the
same energy as in the dVK picture.  However this energy will not
be in the form of a beam, as argued by dVK, but rather, will be 
present in
an isotropic distribution.  Then the IC process still works,
and one of the points made by dVK, that much of the energy is
lost by IC hard X-rays instead of `soft' X-rays from material
evaporated from the disk, can remain valid, as already
noted at the end of \S 1.  In order to see
if the inverse Compton losses actually dominate,
detailed computations of these effects should be undertaken
under various circumstances.

It is well known that in the case of the solar corona, the directly
accelerated beams should be thermalized within a very short time through BPI
(e.g., Sturrock \cite{sturrock}).  In the standard picture, this produces Langmuir waves 
which then manifest themselves as various types
of radio bursts if non-linear effects or transport from faster to
slower electrons within the beam could dominate (e.g., Vlahos \& Raoult
\cite{vlahos}).  
However, energetic electrons have been observed in 
satellite measurements in near-earth orbit, and the outstanding question
of the maintenance of these beams through their propagation from the
sun to the earth has given rise to more complex models involving
complex profiles of the electron beams (Vlahos \& Raoult \cite{vlahos}).  
Instead of producing
X-ray flares via a primary process as proposed by dVK, 
these secondary processes involving energy input to the plasma could
contribute to  variability in the radio band. 

\begin{acknowledgements}  
We thank the anonymous referee for pointing out the incompleteness
of our analysis in the original version.
This work was supported in part by NASA
grant NAG 5-3098 and RPI and Strategic International Initiative 
funds at GSU.
\end{acknowledgements}


\begin{thebibliography}{}

\bibitem[1993]{benz} Benz, A.O., 1993, Plasma Astrophysics.  
Kluwer, Dordrecht

\bibitem[1992]{deVries} de Vries, M.,  Kuijpers, J., 
1992, A\&A 266, 77 (dVK)

\bibitem[1979]{galeev} Galeev, A.A., Rosner, R., Vaiana, G.S., 
1979, ApJ 229, 318

\bibitem[1999]{krishan} Krishan, V., 1999, Astrophysical Plasmas 
and Fluids. Kluwer, Dordrecht 

\bibitem[1995]{kuijpers} Kuijpers, J., 1995, in: Benz, A.O., Kr\"uger, A. (eds.) Coronal
	Magnetic Energy Releases. Springer, Berlin, p. 135

\bibitem[1985]{kuperus} Kuperus, M., Ionson, J.A., 1985, A\&A 148, 309

\bibitem[1977]{liang} Liang, E.P., Price, R.H., 1977, ApJ, 218, 247

\bibitem[1986]{melrose} Melrose, D.B., 1986, Instabilities in Space and Laboratory
Plasmas.  Cambridge University Press, Cambridge, p.\ 72

\bibitem[1961]{rose} Rose, D.J., Clark Jr., M., 1961,
Plasmas and Controlled
Fusion (Massachusetts Institute of Technology Press, Cambridge), p.\ 448

\bibitem[1974]{mikhailovskii} Mikhailovskii, A.B., 1974, Theory of Plasma Instabilities, Vol. I.
	Consultants Bureau, New York

\bibitem[1973]{shakura} Shakura, N.I., Sunyaev, R.A., 1973, A\&A 24, 337

\bibitem[1976]{shapiro} Shapiro, S.L., Lightman, A.P., Eardley, E.M., 1976, ApJ 204, 187

\bibitem[1964]{sturrock} Sturrock, P.A., 1964, in: Hess, W.N. (ed.) Proc. AAS--NASA Symp. on the
	Physics of Solar Flares (NASA SP-50), p. 357

\bibitem[1997]{treumann} Treumann, R.A., Baumjohann, W., 1997, Advanced Space Plasma Physics.
	Imperial College Press, London, p.\ 193

\bibitem[1995]{vlahos} Vlahos, L., Raoult, A., 1995, A\&A, 296, 844

\bibitem[1980] {walsh} Walsh, J.E.  1980, in: Jacobs, S.F. et al. (eds.) Free-Electron
	Generators of Coherent Radiation. Addison-Wesley, Reading, p.\ 255 

\end{thebibliography}
\end{document}